# Quantization of a Deep-Subwavelength-Aperture-Confined Optical Near Field

Hongliang Dang,[1] Jiaxin Gao,[1] Yudong Zhang,[1] Bo Yan,[2] Xin Guo,[1] Pan Wang,[1] Jianqi Shen,[1*] and Limin Tong[1†]

[1]*New Cornerstone Science Laboratory, State Key Laboratory of Extreme Photonics and Instrumentation, College of Optical Science and Engineering, Zhejiang University, Hangzhou 310027, Zhejiang, China*

[2]*Zhejiang Key Laboratory of Micro-nano Quantum Chips and Quantum Control, School of Physics, Zhejiang University, Hangzhou 310027, Zhejiang, China*

[*]jqshen@zju.edu.cn

[†]phytong@zju.edu.cn

# ABSTRACT

Deep-subwavelength apertures are the most elementary structures for confining light far below the free-space wavelength, yet their near fields, which are dominated by evanescent components, have usually been described classically or treated as corrections to propagating-mode quantization. Here we develop a quantized description of the optical near field confined by a deep-subwavelength circular aperture in a perfectly conducting thin screen, starting from Bethe's effective magnetic- and electric- dipole model. Working in the Coulomb gauge, we couple the quantized incident field linearly to the operator-valued Bethe moments, solve the Heisenberg equations with adiabatic switch-on to obtain the steady-state coherent amplitude, and express the resulting vector potential in the Weyl angular-spectrum representation. The finite aperture size enters through an analytical spherical-Bessel form factor, $3j_1(\boldsymbol{k}_\perp a)/(\boldsymbol{k}_\perp a)$, so that propagating ($\boldsymbol{k}_\parallel < \boldsymbol{k}_0$) and evanescent ($\boldsymbol{k}_\parallel > \boldsymbol{k}_0$) components appear in a single expression, while the longitudinal Coulomb field of the Bethe polarization is retained. For coherent monochromatic excitation, the field expectation values reduce to classical dipole radiation in the far field and recover Bethe's transmission cross section, $\sigma \propto k_0^4 a^6$. The framework provides an operator-based basis for nonclassical excitation, quantum fluctuations, and light-matter interactions in deep-subwavelength-confined optical near fields.

# I. INTRODUCTION

The confinement of electromagnetic fields by deep-subwavelength structures, which enables the probing and manipulation of matter on scales far below the free-space optical wavelength, is the base of modern micro- and nano-optics. Across all deep-subwavelength confinement structures, either dielectric [1–3] or plasmonic [4–6], the small aperture is the most representative structure, providing a universal basis for comprehending more complex configurations. Fundamentally, optical transmission through an aperture can be conceptualized as the interaction between magnetic and electric dipoles and a localized electromagnetic field. In 1944, Bethe pioneered this physical framework to analyze diffraction of deep-subwavelength apertures [7]. He demonstrated that for a deep-subwavelength circular aperture (radius $a \ll \lambda_0$, where $\lambda_0$ is the free-space wavelength) in a perfectly conducting thin screen, the diffracted field is equivalent to the superposition of radiation from an effective magnetic dipole moment $\boldsymbol{m_0}$ and an effective electric dipole moment $\boldsymbol{p_0}$ located at the aperture center [7]. Bethe's framework, while adequate for a classical near-to-far-field description of the aperture, is insufficient for describing near-field quantum properties such as quantum fluctuations or noise. Recent experiments in extreme optics have successfully confined light down to sub-1-nm spatial scale in both metallic [8–11] and dielectric nanostructure [12–15], or tens of attoseconds in time scale [16,17]. On such an extreme spatiotemporal scale, the field's quantum fluctuations or noise fundamentally affect observable electron dynamics [18–20] and uncertainty relations for photons [21–23] and thus inevitably demands an appropriate quantized description of such ultra-confined optical near fields.

Over the past half century, the quantization of optical near fields has been continuously carried out. However, due to the relatively loose optical confinement in spatial scale, in these works, the fields are typically quantized based on free-space-like propagating modes (i.e., real wavevector $\boldsymbol{k}$), while localized evanescent components (i.e., imaginary wavevector $\boldsymbol{k}$) entering as boundary-induced or scatterer-induced corrections when deemed necessary. This hierarchy was well suited to the macroscopic and mesoscopic settings in which near-field quantization was originally developed: free-space fields were quantized in homogeneous plane waves, evanescent waves at the infinite homogeneous interface were embedded into complete scattering-mode triplets, and dielectric or absorbing media were treated through Maxwell eigenmodes, polariton operators, or Green-function expansions [24–27]. More recent QED theories of small apertures further extended this propagating-mode-based framework to diffraction, tunneling, and scattering processes that connect incident quantum fields to outgoing channels [28,29]. However, deep-subwavelength confinement changes the role of the evanescent sector. For a deep-subwavelength-aperture-confined field, especially at sub-1-nm spatial scales, high-transverse-wavevector components are no longer secondary corrections to a propagating optical mode; they constitute the localized near-field mode itself. Quantum studies have meanwhile demonstrated the relevance of nonclassical states and fluctuation-sensitive observables in deep-subwavelength-confined fields, including the SPASER proposal for coherent surface-plasmon generation [30], transmission of photon entanglement through a subwavelength hole array [31], propagation of quadrature-squeezed plasmons [32], and squeezed-vacuum-driven photoemission from a nanometric metal tip [33]. These studies show that strongly confined fields require an operator-based description to capture nonclassical input states, quantum observables, and both propagating and evanescent components.

Here, we develop such a quantization framework for deep-subwavelength-confined optical near fields based on the Bethe-equivalent dipole model. Starting from the Bethe electric and magnetic moments, we use the standard transverse photon basis to construct the source-driven field and express it in Weyl representation. This formulation retains the propagating and evanescent spatial components within a single angular-spectrum expression, while the finite aperture size enters through an analytical Bessel form factor. The resulting quantized vector-potential operator therefore provides a unified description of aperture-confined optical fields incorporating both the Bethe-equivalent source response and the finite aperture geometry. For monochromatic coherent excitation, the resulting field expectation values recover the classical dipole radiation field and Bethe's transmission cross section.

# II. DIPOLE MODEL

In the classical framework, compared with other approaches, Bethe's effective-dipole model is more appropriate for scaling down to sub-nm level and describing a deep-subwavelength-confined optical field. Therefore, for quantization, we adopt the equivalent dipole model as the starting point. As conceptually illustrated in Fig. 1(a), the diffraction of an incident single-frequency plane wave after penetrating a deep-subwavelength aperture generates a localized optical near field that exhibits rapid spatial decay. Following Bethe's formulation, this spatially confined field can be rigorously mapped onto the coherent superposition of radiation from an effective magnetic dipole $\boldsymbol{m}(t)$ [Fig. 1(b), where the green arrow along $-x$ denotes the orientation of the magnetic dipole moment] and an effective electric dipole $\boldsymbol{p}(t)$ [Fig. 1(c), where the blue arrow along $+z$ denotes the orientation of the electric dipole moment] localized at the coordinate origin, both harmonically oscillating at a frequency of $\omega_0$, as follows:

$$\boldsymbol{m}(t) = \boldsymbol{m_0} e^{-i\omega_0 t} + c.c., \tag{1}$$

$$\boldsymbol{p}(t) = \boldsymbol{p_0} e^{-i\omega_0 t} + c.c., \tag{2}$$

in which $\boldsymbol{m}_0$ and $\boldsymbol{p}_0$ are the complex amplitudes of the magnetic and electric dipole moments, and can be obtained as:

$$\boldsymbol{m}_0 = -\frac{8}{3} a^3 \boldsymbol{H}_0, \tag{3}$$

$$\boldsymbol{p}_0 = \frac{4}{3} \epsilon_0 a^3 \boldsymbol{E_0}, \tag{4}$$

where $a$ is the radius of the circular aperture, $\epsilon_0$ is the vacuum dielectric constant, $\boldsymbol{H}_0$ and $\boldsymbol{E}_0$ are the incident magnetic and electric field complex amplitudes at the aperture plane, respectively.

Then, the radiated dipolar fields (i.e., the near and far fields after the aperture), can be described in forms of vector $\boldsymbol{A}(\boldsymbol{r}, t)$ and scalar potential $\varphi(\boldsymbol{r}, t)$, and the vector potential satisfies the classical Helmholtz wave equation:

$$\nabla^2 \boldsymbol{A} - \frac{1}{c^2} \frac{\partial^2 \boldsymbol{A}}{\partial t^2} = -\mu_0 (\boldsymbol{J}_\mathrm{m} + \boldsymbol{J}_\mathrm{p}), \tag{5}$$

in which $c$ is the speed of light, $\mu_0$ is the permeability in vacuum and the magnetization current density $\boldsymbol{J}_\mathrm{m}$ and the electric polarization current density $\boldsymbol{J}_\mathrm{p}$ can be obtained as:

$$\boldsymbol{J}_\mathrm{m} = \nabla \times \boldsymbol{M}(t), \tag{6}$$

$$\boldsymbol{J}_\mathrm{p} = \frac{\partial \boldsymbol{P}(t)}{\partial t}, \tag{7}$$

where the magnetic or electric dipole polarization density $\boldsymbol{M}(\boldsymbol{r}, t)$ or $\boldsymbol{P}(\boldsymbol{r}, t)$, is obtained by the average dipole moment over its corresponding volume $V$:

$$\boldsymbol{M}(\boldsymbol{r}, t) = \frac{<\boldsymbol{m}(t)>}{V}, \tag{8}$$

$$\boldsymbol{P}(\boldsymbol{r}, t) = \frac{<\boldsymbol{p}(t)>}{V}. \tag{9}$$

The scalar potential satisfies the Poisson equation:

$$\nabla^2 \varphi(\boldsymbol{r}, t) = -\frac{\rho(r,t)}{\epsilon_0}, \tag{10}$$

where $\rho(\boldsymbol{r},t) = -\nabla \cdot \boldsymbol{P}(\boldsymbol{r},t)$ is the effective polarization charge density determined by the electric dipole polarization density, and hence the scalar potential can be solved as:

$$\varphi(\boldsymbol{r},t) = \frac{1}{4\pi\epsilon_0}\int dr' \frac{\rho(\boldsymbol{r}',t)}{|r-\boldsymbol{r}'|}. \tag{11}$$

Finally, the classical expressions of the radiated magnetic and electric fields of the aperture can be obtained by

$$\boldsymbol{B}(\boldsymbol{r},t) = \nabla \times \boldsymbol{A}(\boldsymbol{r},t), \tag{12}$$

$$\boldsymbol{E}(\boldsymbol{r},t) = -\nabla\varphi(\boldsymbol{r},t) - \frac{\partial \boldsymbol{A}(\boldsymbol{r},t)}{\partial t}, \tag{13}$$

with $\boldsymbol{A}(\boldsymbol{r},t)$ being a function of $\boldsymbol{M}(\boldsymbol{r},t)$ and $\boldsymbol{P}(\boldsymbol{r},t)$, and $\varphi(\boldsymbol{r},t)$ completely determined by the incident optical fields. The preceding construction separates the aperture field into an independent transverse part and a constrained longitudinal part. Upon quantization, only the transverse vector potential is treated as an independent canonical field, whereas the scalar potential is determined by the aperture charge density, $\rho = -\nabla \cdot \boldsymbol{P}$. For a quantum incident field, the Bethe polarization and hence $\rho$, $\varphi$, and $-\nabla\varphi$ become operator-valued. The complete quantum electric field therefore contains both contributions, $\hat{\boldsymbol{E}} = -\dot{\boldsymbol{A}} - \nabla\hat{\varphi}$.

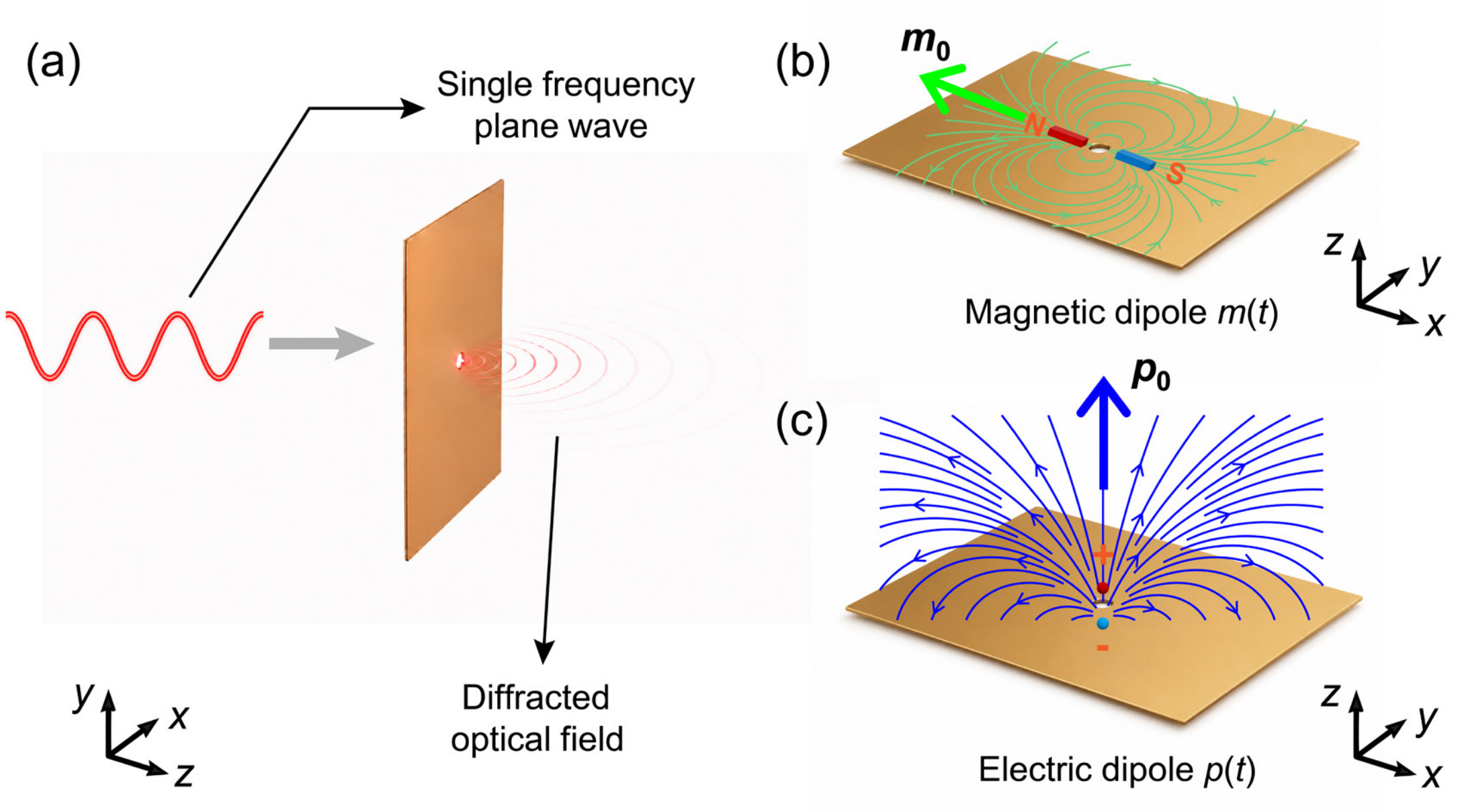

**FIG. 1.** Physical equivalence of aperture diffraction based on Bethe's dipole model. **(a)** A single-frequency plane wave impinges on a deep-subwavelength aperture, generating a localized diffracted optical field. Following Bethe's framework, this localized near field can be rigorously formulated as the coherent superposition of radiation from **(b)** an effective magnetic dipole oriented parallel to the plane and **(c)** an effective electric dipole oriented perpendicular to the aperture plane, both localized at the aperture center.

# III. SOURCE-DRIVEN QUANTUM APERTURE FIELD IN THE WEYL REPRESENTATION

Based on the Bethe-equivalent aperture sources and the transverse–longitudinal structure of their fields, we now formulate this source-driven description on a canonical basis. As detailed in the Supporting Information, canonical quantization in the Coulomb gauge transfers the transverse vector potential to the field operator denoted by the wave vector $\boldsymbol{k}$ and polarization index $\lambda$:

$$\widehat{\boldsymbol{A}}(\boldsymbol{r},t) = N_{\mathbf{k}} \cdot \int d^3k \sum_{\lambda=1,2} \left[ \hat{a}_{\mathbf{k}\lambda}\, \boldsymbol{\varepsilon}_{\mathbf{k}\lambda}\, e^{-i(\omega_{\mathbf{k}}t - \boldsymbol{k}\cdot\boldsymbol{r})} + \mathrm{h.c.} \right], \tag{14}$$

where $\omega_{\mathbf{k}} = c|\boldsymbol{k}|$, $\boldsymbol{\varepsilon}_{\mathbf{k}\lambda}$ are the polarization unit vectors orthogonal to $\boldsymbol{k}$, $\hat{a}_{\mathbf{k}\lambda}$ is the annihilation operator, h.c. is the Hermitian conjugate, and $N_{\mathbf{k}}$ is the normalization coefficient determined as follows:

$$N_{\mathbf{k}} = \sqrt{\frac{\hbar}{2\epsilon_0\, \omega_{\mathbf{k}}\, (2\pi)^3}}. \tag{15}$$

It's worth mentioning that, under the Gupta–Bleuler condition, the Lorenz- and Coulomb-gauge formulations yield the same physical electromagnetic fields [see more details in S1]. Therefore, for better clarity, we adopt the Coulomb-gauge transverse operator as the canonical basis in what follows.

Using this canonical basis, we now couple the quantum incident field (with an oscillating frequency of $\omega_0$ and a free-space wavenumber of $k_0 = \omega_0/c$) to the Bethe-equivalent electric and magnetic moments and construct the resulting monochromatic aperture field. Because the aperture is localized in the transverse plane, the resulting field is naturally represented by the Weyl angular spectrum, which separates its propagating components (i.e., $|\boldsymbol{k}_\perp| < k_0$ and $\boldsymbol{k}_{\mathbf{z}} = \sqrt{k_0^2 - \boldsymbol{k}_\perp^2}$, where $k_\perp$ is the transverse wavevector) from its evanescent components (i.e., $|\boldsymbol{k}_\perp| > k_0$ and $\boldsymbol{k}_{\mathbf{z}} = i\kappa = i\sqrt{\boldsymbol{k}_\perp^2 - k_0^2}$). The latter constitute an essential part of the deep-subwavelength-aperture confined near field.

## Interaction Hamiltonian

For an incident monochromatic mode $(\boldsymbol{k}_0, \lambda_0)$, at the aperture the electric and magnetic fields are thus written as

$$\widehat{E}_{\mathrm{in}}(t) = E_{\mathrm{N}} \hat{a}_{\mathrm{in}} e^{-i(\omega_0 t - \boldsymbol{k}_0 \boldsymbol{r})} + \mathrm{h.c.},\ \widehat{H}_{\mathrm{in}}(t) = H_{\mathrm{N}} \hat{a}_{\mathrm{in}} e^{-i(\omega_0 t - \boldsymbol{k}_0 \boldsymbol{r})} + \mathrm{h.c.}, \tag{16}$$

where $\hat{a}_{\rm in}$ is the annihilation operator of the incident mode, and the normalized coefficients are written as

$$E_{\rm N} = i\omega_0 N_0 \boldsymbol{\varepsilon_0},\ \ H_{\rm N} = i\mu_0 N_0 (\boldsymbol{k_0} \times \boldsymbol{\varepsilon_0}), \tag{17}$$

with $N_0 = \sqrt{\frac{\hbar}{2\epsilon_0\,\omega_0\,(2\pi)^3}}$ the canonical normalization coefficient of the incident mode and $\boldsymbol{\varepsilon_0}$ the polarization unit vector. For a single-photon incident state $|\,1_{\rm in}\rangle$,

$$\hat{a}_{\rm in}\,|\,1_{\rm in}\rangle = \sqrt{1}\,|\,0_{\rm in}\rangle. \tag{18}$$

Since the Bethe response is linear in the incident field, Eqs. (1) and (2) can be extended directly to the quantized incident mode by replacing the classical field amplitudes with the corresponding field operators written as:

$$\widehat{\boldsymbol{m}}(t) = -\frac{8}{3}a^3 H_{\rm N}\hat{a}_{\rm in}e^{-i\omega_0 t} + \text{h.c.}, \tag{19}$$

$$\widehat{\boldsymbol{p}}(t) = \frac{4}{3}a^3 E_{\rm N}\hat{a}_{\rm in}e^{-i\omega_0 t} + \text{h.c.}, \tag{20}$$

with $-\frac{8}{3}a^3 H_{\rm N} = \boldsymbol{m}_{\rm N}$ and $\frac{4}{3}a^3 E_{\rm N} = \boldsymbol{p}_{\bf N}$. The induced dipole moments then couple to the quantized electromagnetic field through the standard electric- and magnetic-dipole interaction:

$$\widehat{H}_{\rm int}(t) = -\widehat{\boldsymbol{m}}(t)\cdot\left(\nabla\times\widehat{\boldsymbol{A}}\right)|_{\boldsymbol{r}=0} + \widehat{\boldsymbol{p}}(t)\cdot\frac{\partial\widehat{\boldsymbol{A}}}{\partial t}|_{\boldsymbol{r}=0}. \tag{21}$$

The corresponding mode-coupling matrix element for scattering one quantum from the incident mode into the aperture confined mode $(\boldsymbol{k},\lambda)$ is:

$$V_{\rm S}(\boldsymbol{k},\boldsymbol{\lambda}) = \left\langle 1_{\bf k\lambda}, 0_{\rm in}\middle|\widehat{H}_{\rm int}\middle|0, 1_{\rm in}\right\rangle = -i\,N_{\bf k}\,\boldsymbol{\varepsilon}^*_{\bf k\lambda}\cdot(\boldsymbol{k}\times\boldsymbol{m}_{\rm N} - \omega_{\bf k}\boldsymbol{p}_{\rm N\perp}) \tag{22}$$

where the transverse component of the electric dipole moment is $\boldsymbol{p}_{\rm N\perp} = \boldsymbol{n}\times(\boldsymbol{p}_{\rm N}\times\boldsymbol{n})$, and $\boldsymbol{n} = \boldsymbol{k}/|\,\boldsymbol{k}\,|$ is the unit vector along the radiation wave vector.

## Heisenberg Evolution and Coherent-State Amplitude

Under continuous monochromatic driving, the aperture source builds a steady-state coherent field. To obtain the coherent-state amplitude, we solve the Heisenberg equation with adiabatic switch-on:

$$i\hbar\frac{d}{dt}\hat{a}_{\bf k}(t) = \left[\hat{a}_{\bf k}(t), \widehat{H}\right], \tag{23}$$

where the total Hamiltonian operator is:

$$\hat{H} = \hat{H}_0 + \hat{H}_{\rm int} = \hbar\omega_{\rm k}\hat{a}^{\dagger}_{\mathbf{k}}\hat{a}_{\mathbf{k}} + V_{\rm S}(\boldsymbol{k},\boldsymbol{\lambda})\hat{a}^{\dagger}_{\mathbf{k}}\hat{a}_{\mathbf{in}} + V^{*}_{\rm S}(\boldsymbol{k},\boldsymbol{\lambda})\hat{a}^{\dagger}_{\mathbf{in}}\hat{a}_{\mathbf{k}}. \tag{24}$$

After simplification, the Heisenberg equation of motion can be written as:

$$\frac{d}{dt}\left(\hat{a}_{\mathbf{k}}(t)e^{i\omega_{\mathbf{k}}t}\right) = -\frac{i}{\hbar}V_{\rm S}(\boldsymbol{k},\boldsymbol{\lambda})\cdot\hat{a}_{\rm in}\cdot e^{i(\omega_k-\omega_0-i\eta)t}, \tag{25}$$

where $\eta \to 0^{+}$ is the adiabatic switch-on parameter for first-order perturbation theory. Solving Eq. (25) yields:

$$\hat{a}_{\mathbf{k}}(t) = -\frac{V_{\rm S}(\boldsymbol{k},\boldsymbol{\lambda})}{\hbar}\hat{a}_{\rm in}\frac{e^{-i\omega_0 t+\eta t}}{\omega_{\mathbf{k}}-\omega_0-i\eta}. \tag{26}$$

Taking the expectation value of Eq. (26) yields the steady-state coherent-state amplitude:

$$\alpha_{3D}(\boldsymbol{k},\lambda) =< \hat{a}_{\mathbf{k}}(t) >= -\frac{V_{\rm S}(\boldsymbol{k},\boldsymbol{\lambda})}{\hbar}\frac{\alpha_{\rm in}}{\omega_{\mathbf{k}}-\omega_0-i\eta}. \tag{27}$$

Using $E_{\rm N}\alpha_{\rm in} = E_0; H_{\rm N}\alpha_{\rm in} = H_0$, we obtain:

$$V_{3D}(\boldsymbol{k},\boldsymbol{\lambda}) = V_{\rm S}(\boldsymbol{k},\boldsymbol{\lambda})\alpha_{\rm in} = -i\,N_{\mathbf{k}}\,\boldsymbol{\varepsilon}^{*}_{\mathbf{k}\lambda}\cdot(\boldsymbol{k}\times\boldsymbol{m}_0 - \omega_{\mathbf{k}}\boldsymbol{p}_{\mathbf{0}\perp}). \tag{28}$$

In the three-dimensional coherent state $|\{\alpha_{3D}(\boldsymbol{k},\lambda)\}\rangle$, the expectation value of the vector-potential operator is:

$$\langle\hat{\boldsymbol{A}}(\boldsymbol{r},t)\rangle = \int d^2k_{\perp}\int dk_{\rm z}\cdot\alpha_{3D}(\mathbf{k},\lambda)\cdot N_{\mathbf{k}}\,\boldsymbol{\varepsilon}_{k_{\perp}\lambda}\,e^{-i(\omega_{\mathbf{k}}t-\boldsymbol{k}\cdot\boldsymbol{r})}. \tag{29}$$

Carrying out the change of variables using the Jacobian determinant $dk_{\rm z} = d\omega_{\rm k}\frac{\omega_{\mathbf{k}}}{c^2k_{\mathbf{z}}}$ yields:

$$\langle\hat{\boldsymbol{A}}(\boldsymbol{r},t)\rangle = \int d^2k_{\perp}\cdot -\frac{2\pi i\omega_0}{\hbar c^2k_z}\cdot N_{\mathbf{k}}\cdot V_{3D}(\mathbf{k},\lambda)\cdot\,\boldsymbol{\varepsilon}_{k_{\perp}\lambda}\,e^{-i(\omega_0 t-\boldsymbol{k}\cdot\boldsymbol{r})}. \tag{30}$$

This relation converts the canonical three-dimensional mode normalization into the two-dimensional Weyl angular spectrum used below.

# IV. RESULTS

To obtain the magnetic vector potential for a circular aperture of finite radius, we incorporate the spatial distribution of the Bethe-equivalent sources into the Weyl representation developed above. The aperture geometry enters the angular spectrum through an analytical form factor. The resulting expression for the coherent-state expectation value retains both propagating and evanescent contributions. We then evaluate its far-field limit to verify consistency with Bethe's dipole field and transmission cross section.

## Form Factor of a Circular Aperture

An actual aperture has finite transverse extent. To retain this geometry, consider an infinitely thin, perfectly conducting screen occupying the plane $z = 0$, with a circular aperture of radius $a$ centered at the origin. The conducting screen occupies the region $\boldsymbol{\varrho} > a$ (where $\boldsymbol{\varrho} = \sqrt{\boldsymbol{x}^2 + \boldsymbol{y}^2}$), and the aperture covers $\boldsymbol{\varrho} \leq a$.

When a finite circular aperture of radius *a* is considered, the induced current distribution in the aperture plane is spatially broadened. Bethe gave the magnetic- and electric- dipole densities within the aperture as follows:

$$\widetilde{\boldsymbol{m}}(\boldsymbol{\varrho}) = \boldsymbol{m_0} f(\boldsymbol{\varrho}), \quad \boldsymbol{\varrho} \leq a, \tag{31}$$

$$\widetilde{\boldsymbol{p}}(\boldsymbol{\varrho}) = \boldsymbol{p_{0\perp}} f(\boldsymbol{\varrho}), \quad \boldsymbol{\varrho} \leq a, \tag{32}$$

where the aperture form factor $f(\boldsymbol{\varrho})$ is given by:

$$f(\boldsymbol{\varrho}) = \frac{3}{2\pi a^2}\sqrt{1 - \left(\frac{\varrho}{a}\right)^2}, \quad \boldsymbol{\varrho} \leq a \tag{33}$$

The angular spectrum (two-dimensional Fourier transform) of the form factor is:

$$f(\boldsymbol{k}_\perp) = \int d^2\varrho \, f(\boldsymbol{\varrho}) \, e^{-i\boldsymbol{k}_\perp \cdot \boldsymbol{\varrho}} . \tag{34}$$

Transforming Eq. (34) to the polar coordinates $(\boldsymbol{\varrho}, \phi)$, and exploiting azimuthal symmetry together with the definition of the zeroth-order cylindrical Bessel function $\int_0^{2\pi} e^{-ik_\perp \varrho \cos\phi} \, d\phi = 2\pi J_0(\boldsymbol{k}_\perp \boldsymbol{\varrho})$, we obtain:

$$f(\boldsymbol{k}_\perp) = 2\pi \int_0^a \boldsymbol{\varrho} \, d\varrho \, \frac{3}{2\pi a^2}\sqrt{1 - \left(\frac{\varrho}{a}\right)^2} \, J_0(\boldsymbol{k}_\perp \boldsymbol{\varrho}) . \tag{35}$$

Setting $\boldsymbol{u} = \boldsymbol{\varrho}/a$:

$$f(\boldsymbol{k}_\perp) = 3 \int_0^1 \boldsymbol{u} \, du \sqrt{1 - \boldsymbol{u}^2} J_0(\boldsymbol{k}_\perp a \, \boldsymbol{u}) . \tag{36}$$

This integral is given by Sonine's finite integral [34]

$$\int_0^1 \boldsymbol{u} \sqrt{1 - \boldsymbol{u}^2} J_0(c \, \boldsymbol{u}) \, du = \frac{j_1(c)}{c}, \tag{37}$$

This is the special case when $\nu = \frac{1}{2}$ *and* $\mu = 0$, of the general formula $\int_0^1 \boldsymbol{u}^{\mu+1}(1 - \boldsymbol{u}^2)^\nu J_\mu(c\boldsymbol{u})\, du = 2^\nu \Gamma(\nu+1) c^{-(\nu+1)} J_{\mu+\nu+1}(c)$. Using the relation between cylindrical and spherical Bessel functions, $j_1(c) = \sqrt{\frac{\pi}{2c}} J_{\frac{3}{2}}(c)$ and $\Gamma(3/2) = \sqrt{\pi}/2$, we arrive at:

$$f(\boldsymbol{k}_\perp) = 3\,\frac{j_1(\boldsymbol{k}_\perp a)}{\boldsymbol{k}_\perp a}, \tag{38}$$

where $j_1$ is the spherical Bessel function of the first kind of order one. The corresponding angular spectrum of magnetic- and electric- dipole densities are obtained as:

$$\widetilde{\boldsymbol{m}}(\boldsymbol{k}_\perp) = \boldsymbol{m}_{\mathbf{0}} 3\,\frac{j_1(\boldsymbol{k}_\perp a)}{\boldsymbol{k}_\perp a}, \tag{39}$$

$$\widetilde{\boldsymbol{p}}(\boldsymbol{k}_\perp) = \boldsymbol{p}_{\mathbf{0}\perp} 3\,\frac{j_1(\boldsymbol{k}_\perp a)}{\boldsymbol{k}_\perp a}, \tag{40}$$

## Complete Result Incorporating the Aperture Form Factor

Combining the source-driven coherent amplitude with the magnetic- and electric-dipole densities gives:

$$V_{k_\perp,\lambda}^{(\text{hole})} = -i\, N_{k_\perp}\, \boldsymbol{\varepsilon}_{k_\perp\lambda}^{*} \cdot \left(\boldsymbol{k} \times \widetilde{\mathbf{m}}(\boldsymbol{k}_\perp) - \omega_0 \widetilde{\boldsymbol{p}}(\boldsymbol{k}_\perp)\right). \tag{41}$$

The complete result for the expectation value of the vector potential in the right half-space is:

$$\langle \widehat{\boldsymbol{A}}(\boldsymbol{r},t) \rangle = -\frac{\mu_0}{8\pi^2} \int d^2 k_\perp\, \frac{\boldsymbol{k} \times \boldsymbol{m}_0 - \omega_0 \boldsymbol{p}_{\mathbf{0}\perp}}{\boldsymbol{k}_{\mathbf{z}}} \cdot \frac{3 j_1(\boldsymbol{k}_\perp a)}{\boldsymbol{k}_\perp a}\, e^{-i(\omega_0 t - \mathbf{k}_\perp \cdot \varrho - k_z z)}. \tag{42}$$

This result unifies the Bethe magnetic- and electric-dipole response with the finite aperture geometry in a single angular-spectrum expression. It applies from the near-field region ($\boldsymbol{z} \lesssim a$), where evanescent components dominate, to the far field ($\boldsymbol{k}_\perp a \ll 1$, $j_1(\boldsymbol{k}_\perp a) \approx \frac{1}{3} \boldsymbol{k}_\perp a$), where the expression reduces to the classical Bethe limit.

Equation (42) gives the transverse vector-potential expectation value. The corresponding observable field amplitudes are obtained as：

$$< \widehat{\boldsymbol{B}}(\boldsymbol{r},t) >= \nabla \times < \widehat{\boldsymbol{A}}(\boldsymbol{r},t) >, \tag{43}$$

$$< \widehat{\boldsymbol{E}}(\boldsymbol{r},t) >= i\omega_0 \cdot < \widehat{\boldsymbol{A}}(\boldsymbol{r},t) > -\nabla < \widehat{\varphi}(\boldsymbol{r},t) >, \tag{44}$$

where $< \hat{\varphi}(\boldsymbol{r}, t) >$ is determined by the polarization charge density and can be described as follows:

$$< \hat{\varphi}(\boldsymbol{r}, t) >= \frac{1}{4\pi\epsilon_0} \int d^3r' \frac{<\hat{\rho}(\boldsymbol{r}', t)>}{|\boldsymbol{r}-\boldsymbol{r}'|}. \tag{45}$$

Thus, the complete field expectation values retain both the transverse radiation field and the longitudinal Coulomb field generated by the Bethe polarization, without introducing an independent scalar-field oscillator.

## Far-field Limit: Recovery of the Bethe Results

In the far-field limit, the Bessel form factor in Eq. (42) tends to unity and thus the vector potential reduces to:

$$\langle \hat{\boldsymbol{A}}(\boldsymbol{r}, t) \rangle = -\frac{\mu_0}{8\pi^2} \int d^2k_\perp \frac{\boldsymbol{k} \times \boldsymbol{m}_0 - \omega_0 \boldsymbol{p}_{\mathbf{0}\perp}}{k_{\mathbf{z}}} e^{-i(\omega_0 t - \mathbf{k}_\perp \cdot \boldsymbol{\varrho} - k_z z)}. \tag{46}$$

Applying the Sommerfeld identity [35] to Eq. (46) yields:

$$\langle \hat{\boldsymbol{A}}(\boldsymbol{r}, t) \rangle = \frac{\mu_0}{4\pi} \nabla \times \left( \boldsymbol{m}_0 \frac{e^{-i(\omega_0 t - \boldsymbol{k}_\mathbf{0} r)}}{r} \right) - \frac{i\mu_0\omega_0}{4\pi} \boldsymbol{p}_{\mathbf{0}\perp} \frac{e^{-i(\omega_0 t - \boldsymbol{k}_\mathbf{0} r)}}{r}. \tag{47}$$

The classical value of the Bethe vector potential [36] is obtained by substituting Eqs. (3) and (4) into Eq. (42):

$$\boldsymbol{A} = \frac{i\mu_0 a^3}{3\pi} [(\boldsymbol{k}_\mathbf{0} \times 2\boldsymbol{H}_\mathbf{0}) \cdot \nabla \frac{e^{-i(\omega_0 t - \boldsymbol{k}_\mathbf{0} r)}}{\boldsymbol{r}} + \omega_0 \cdot (\boldsymbol{n} \times \epsilon_0 \boldsymbol{E}_\mathbf{0} \times \boldsymbol{n}) \cdot \frac{e^{-i(\omega_0 t - \boldsymbol{k}_\mathbf{0} r)}}{\boldsymbol{r}}]. \tag{48}$$

From this result, we could further calculate the total radiated power by integrating the corresponding dipole radiation pattern over the transmitted far-field hemisphere under normal incidence condition (i.e., $\boldsymbol{p}_\mathbf{0} = 0$):

$$P_{rad} = \frac{\mu_0 c \boldsymbol{k}_\mathbf{0}^4}{6\pi} (|\boldsymbol{m}_\mathbf{0}|^2 + c^2 \boldsymbol{p}_\mathbf{0}^2), \tag{49}$$

To obtain the transmission cross section, we normalize the radiated power by the time-averaged incident energy flux through the aperture plane:

$$S_{inc} = \frac{1}{2} \mu_0 c \boldsymbol{H}_\mathbf{0}^2 = \frac{1}{2Z_0} \boldsymbol{E}_\mathbf{0}^2. \tag{50}$$

The ratio of the total radiated power to this incident flux therefore yields the cross-section:

$$\sigma_t = \frac{64}{27\pi} k_0^4 a^6, \tag{51}$$

which is the classical Bethe result. The recovery of this scaling is not imposed phenomenologically; it follows from the far-field limit of the quantized vector potential. Thus, the present theory preserves the established far-field aperture response while extending the description to the near-field region where evanescent angular-spectrum components dominate. In this sense, the Bethe cross section provides the far-field projection of the same source-defined quantum aperture mode described by Eq. (42).

# V. DISCUSSION

The recovery of Bethe's transmission cross section confirms that the present formulation reproduces the classical far-field response under coherent excitation. We now examine its relation to Bethe's aperture theory and previous near-field QED formulations to clarify its scope and specific contribution. The discussion focuses on the treatment of the effective aperture sources and the role of the finite-aperture Weyl representation in incorporating aperture geometry into the canonical quantum-field description.

## Relation to Bethe and Previous Near-Field QED

The present construction retains the effective-source content of Bethe's aperture theory: the Bethe polarizabilities encode the conducting-screen response that forms the induced magnetic and electric moments, whereas the microscopic electrons of the screen are not quantized explicitly. Its transverse–longitudinal separation is consistent with Keller's Coulomb-gauge formulation of near-field electrodynamics, in which the transverse vector potential supplies the independent photon degrees of freedom and the longitudinal electric field is constrained by the material charge density. Keller also developed angular-spectrum Green-function descriptions of evanescent fields, while Jung and Keller subsequently formulated small-hole QED in terms of operator-valued multipole moments and propagator-based input–output scattering matrices [28,29,37–39].

The present contribution specializes in the canonical quantum field to a finite Bethe aperture and derives an aperture-centered Weyl representation of the source-generated field. This construction has three practical consequences: it connects the three-dimensional transverse photon basis to the two-dimensional angular spectrum of a localized aperture, retains the aperture geometry through the analytical form factor $3j_1(\boldsymbol{k}_\perp a)/(\boldsymbol{k}_\perp a)$, and provides direct access to spatially resolved electric and magnetic near fields in the region where evanescent components dominate.

# VI. CONCLUSION

Starting from the Bethe dipole model, we present a first-principles quantization of confined optical fields through a deep-subwavelength aperture in both the Coulomb and Lorenz gauges. The equivalence of these two quantization schemes within the physical Hilbert space is rigorously verified via the gauge-invariant field tensor. In the classical limit of coherent states, the quantized magnetic vector potential strictly reduces to the classical Bethe vector potential, recovering the standard Bethe scaling law for the radiation cross section, $\sigma_t = \frac{64}{27\pi} k_0^4 a^6$. For a circular aperture of finite radius, we derive the quantum vector potential (Eq. (42)) incorporating the Bessel form factor, $\frac{3j_1(\boldsymbol{k}_\perp a)}{\boldsymbol{k}_\perp a}$. This expression unifies both the propagating modes ($\boldsymbol{k}_\perp < \boldsymbol{k}_0$) and the evanescent modes ($\boldsymbol{k}_\perp > \boldsymbol{k}_0$), naturally recovering the classical Bethe result in the point-aperture limit ($\boldsymbol{k}_\perp a \ll 1$). In a broader context, because the aperture stands out as the most representative and fundamental deep-subwavelength confinement structure, the unified theoretical framework developed here extends far beyond a specific geometry; it provides a general physical basis for comprehending and quantizing more complex nanophotonic configurations. As modern experiments in extreme optics relentlessly approach the observable limits of quantum effects at the nano-attosecond spatiotemporal scale, the classical approximations inevitably break down. Consequently, a rigorous, complete quantized description of confined optical near fields is no longer merely of formal theoretical interest, but an absolute necessity. By rigorously placing propagating and evanescent modes on an equal footing, this work formally established a robust foundational framework for future explorations of quantum vacuum fluctuations, noise, and highly confined light-matter interactions.

# References

[1] A. I. Kuznetsov, A. E. Miroshnichenko, M. L. Brongersma, Y. S. Kivshar, and B. Luk'yanchuk, Optically resonant dielectric nanostructures, Science **354**, aag2472 (2016).

[2] V. R. Almeida, Q. Xu, C. A. Barrios, and M. Lipson, Guiding and confining light in void nanostructure, Opt. Lett. **29**, 1209 (2004).

[3] K. Chung, Y. Jo, C. Ko, and I. Kim, Extreme light-matter interactions in single-digit nanospaces and their roles in biosensors, chemical processes, and quantum photonics, ACS Nano **19**, 28110 (2025).

[4] W. L. Barnes, A. Dereux, and T. W. Ebbesen, Surface plasmon subwavelength optics, Nature **424**, 824 (2003).

[5] J. A. Schuller, E. S. Barnard, W. Cai, Y. C. Jun, J. S. White, and M. L. Brongersma, Plasmonics for extreme light concentration and manipulation, Nat. Mater. **9**, 193 (2010).

[6] S. Kasani, K. Curtin, and N. Wu, A review of 2D and 3D plasmonic nanostructure array patterns: fabrication, light management and sensing applications, Nanophotonics **8**, 2065 (2019).

[7] H. A. Bethe, Theory of diffraction by small holes, Phys. Rev. **66**, 163 (1944).

[8] F. Benz et al., Single-molecule optomechanics in "picocavities," Science **354**, 726 (2016).

[9] Y. Zhang et al., Visualizing coherent intermolecular dipole–dipole coupling in real space, Nature **531**, 623 (2016).

[10] W. Zhu, R. Esteban, A. G. Borisov, J. J. Baumberg, P. Nordlander, H. J. Lezec, J. Aizpurua, and K. B. Crozier, Quantum mechanical effects in plasmonic structures with subnanometre gaps, Nat. Commun. **7**, 11495 (2016).

[11] Y. Li, W. Chen, X. He, J. Shi, X. Cui, J. Sun, and H. Xu, Boosting light-matter interactions in plasmonic nanogaps, Adv. Mater. **36**, 2405186 (2024).

[12] H. Wu, L. Yang, P. Xu, J. Gong, X. Guo, P. Wang, and L. Tong, Photonic nanolaser with extreme optical field confinement, Phys. Rev. Lett. **129**, 013902 (2022).

[13] L. Yang, Z. Zhou, H. Wu, H. Dang, Y. Yang, J. Gao, X. Guo, P. Wang, and L. Tong, Generating a sub-nanometer-confined optical field in a nanoslit waveguiding mode, Adv. Photonics **5**, 046003 (2023).

[14] Y. Yang, J. Gao, H. Wu, Z. Zhou, L. Yang, X. Guo, P. Wang, and L. Tong, Generating a nanoscale blade-like optical field in a coupled nanofiber pair, Photonics Res. **12**, 154 (2024).

[15] L. Yang et al., Weak-disturbance imaging and characterization of ultra-confined optical near fields, Light: Sci. Appl. **14**, 358 (2025).

[16] P. Eckle, A. N. Pfeiffer, C. Cirelli, A. Staudte, R. Dörner, H. G. Muller, M. Büttiker, and U. Keller, Attosecond ionization and tunneling delay time measurements in helium, Science **322**, 1525 (2008).

[17] A. S. Landsman and U. Keller, Attosecond science and the tunnelling time problem, Phys. Rep. **547**, 1 (2015).

[18] E. Onac, F. Balestro, L. H. W. van Beveren, U. Hartmann, Y. V. Nazarov, and L. P. Kouwenhoven, Using a quantum dot as a high-frequency shot noise detector, Phys. Rev. Lett. **96**, 176601 (2006).

[19] V. Di Giulio and F. J. G. De Abajo, Electron diffraction by vacuum fluctuations, New J. Phys. **22**, 103057 (2020).

[20] A. Feist et al., Cavity-mediated electron-photon pairs, Science **377**, 777 (2022).

[21] I. Bialynicki-Birula, *Photon Wave Function*, arXiv:quant-ph/0508202.

[22] I. Bialynicki-Birula and Z. Bialynicka-Birula, Heisenberg uncertainty relations for photons, Phys. Rev. A **86**, 022118 (2012).

[23] I. Bialynicki-Birula and Z. Bialynicka-Birula, Universal heisenberg uncertainty relations for all massless particles, New J. Phys. **28**, 014506 (2026).

[24] C. K. Carniglia and L. Mandel, Quantization of evanescent electromagnetic waves, Phys. Rev. D **3**, 280 (1971).

[25] R. J. Glauber and M. Lewenstein, Quantum optics of dielectric media, Phys. Rev. A **43**, 467 (1991).

[26] B. Huttner and S. M. Barnett, Quantization of the electromagnetic field in dielectrics, Phys. Rev. A **46**, 4306 (1992).

[27] T. Gruner and D.-G. Welsch, Green-function approach to the radiation-field quantization for homogeneous and inhomogeneous Kramers-Kronig dielectrics, Phys. Rev. A **53**, 1818 (1996).

[28] J. Jung and O. Keller, On optical tunneling in near-field diffraction of light from a small hole, J. Opt. Soc. Am. B **34**, 726 (2017).

[29] J. Jung and O. Keller, QED theory for small-hole diffraction: Young diffraction from two Bethe holes, Phys. Rev. A **104**, 013714 (2021).

[30] D. J. Bergman and M. I. Stockman, Surface plasmon amplification by stimulated emission of radiation: quantum generation of coherent surface plasmons in nanosystems, Phys. Rev. Lett. **90**, 027402 (2003).

[31] E. Altewischer, M. P. van Exter, and J. P. Woerdman, Plasmon-assisted transmission of entangled photons, Nature **418**, 304 (2002).

[32] A. Huck, S. Smolka, P. Lodahl, A. S. Sørensen, A. Boltasseva, J. Janousek, and U. L. Andersen, Demonstration of quadrature-squeezed surface plasmons in a gold waveguide, Phys. Rev. Lett. **102**, 246802 (2009).

[33] J. Heimerl, A. Rasputnyi, J. Pölloth, S. Meier, M. Chekhova, and P. Hommelhoff, Quantum light drives electrons strongly at metal needle tips, Nat. Phys. **21**, 1899 (2025).

[34] G. N. Watson, *A Treatise on the Theory of Bessel Functions* (Cambridge University Press, 1995).

[35] W. C. Chew and S.-Y. Chen, Response of a point source embedded in a layered medium, IEEE Antennas Wireless Propag. Lett. **2**, 254 (2003).

[36] J. D. Jackson, *Classical Electrodynamics, International Adaptation* (John Wiley & Sons, 2021).

[37] O. Keller, On the theory of spatial localization of photons, Phys. Rep., **411**, 1 (2005).

[38] O. Keller, Near-field photon wave mechanics in the Lorenz gauge, Phys. Rev. A **76**, 062110 (2007).

[39] J. Jung and O. Keller, Quantum-mechanical diffraction theory of light from a small hole: extinction-theorem approach, Phys. Rev. A **92**, 012122 (2015).

# SUPPLEMENTAL MATERIAL

# Quantization of a Deep-Subwavelength-Aperture-Confined Optical Near Field

Hongliang Dang,[1] Jiaxin Gao,[1] Yudong Zhang,[1] Bo Yan,[2] Xin Guo,[1] Pan Wang,[1] Jianqi Shen,[1*] and Limin Tong[1†]

[1] *New Cornerstone Science Laboratory, State Key Laboratory of Extreme Photonics and Instrumentation, College of Optical Science and Engineering, Zhejiang University, Hangzhou 310027, Zhejiang, China*

[2] *Zhejiang Key Laboratory of Micro-nano Quantum Chips and Quantum Control, School of Physics, Zhejiang University, Hangzhou 310027, Zhejiang, China*

[*] jqshen@zju.edu.cn

[†] phytong@zju.edu.cn

The main text establishes a quantized description of the optical near field confined by a deep-subwavelength aperture. This Supplemental Material presents the key derivations on which this framework rests: the physical observables are independent of the gauge choice, the operator-valued Bethe source coherently displaces the photon basis, the finite aperture geometry enters the Weyl angular spectrum through the Bessel form factor, and the same quantum field recovers Bethe's radiation law in the far-field limit.

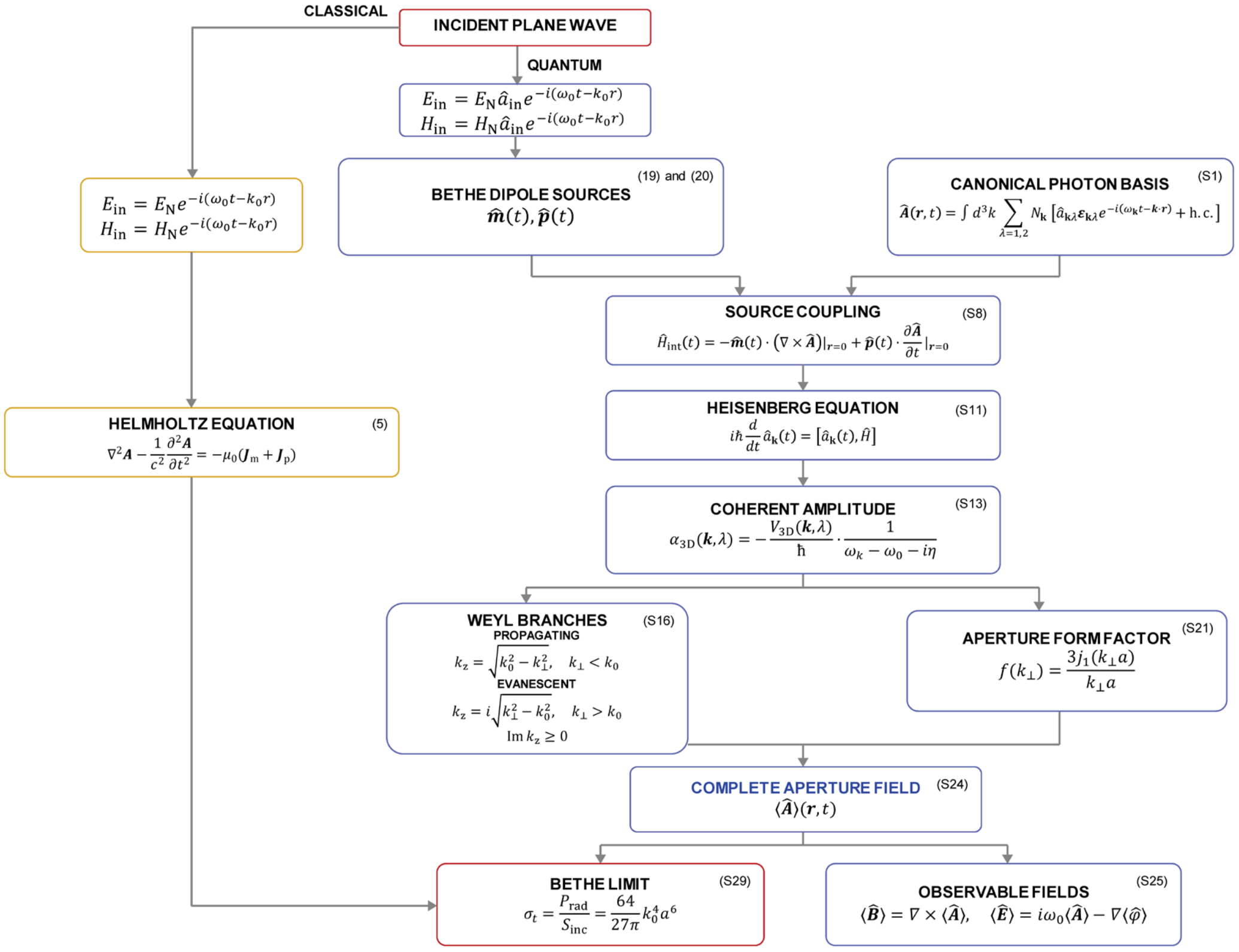

**FIG. S1. Derivation flow for the quantization of a deep-subwavelength-aperture-confined optical near field.** An incident plane wave, treated either as a quantized or as a classical field, induces the Bethe magnetic and electric dipole moments $\hat{m}(t)$ and $\hat{p}(t)$ [Eqs. (19) and (20) in the main text]. Through the interaction Hamiltonian of Eq. (S8), these moments couple to the canonical transverse photon basis of Eq. (S1) and generate the steady-state coherent amplitude $\alpha_{3D}(\boldsymbol{k}, \lambda)$ [Eq. (S13)]. The propagating ($k_\perp < k_0$) and evanescent ($k_\perp > k_0$, Im $k_z > 0$) Weyl branches [Eq. (S16)], weighted by the circular-aperture form factor $f(k_\perp)$ [Eq. (S21)], combine into the complete finite-aperture field [Eq. (S24)]. The observable magnetic and electric fields then follow from Eqs. (S25) and (S26), the latter including the longitudinal contribution of the Bethe polarization charge. In the far-field limit, normalization by the classical incident flux [Eq. (S28)] recovers Bethe's transmission law [Eq. (S29)].

## S1. CANONICAL NORMALIZATION AND GAUGE-INDEPENDENT FIELD CONTENT

To identify the gauge-independent physical content of the source-driven quantization developed in the main text, we establish the equivalence of the Coulomb- and Lorenz-gauge formulations at the level of field observables. We begin with the transverse vector-potential operator labeled by the wave vector $\boldsymbol{k}$ and polarization index $\lambda$ in the Coulomb gauge:

$$\widehat{\boldsymbol{A}}(\boldsymbol{r},t) = \int d^3k \sum_{\lambda=1,2} N_{\mathbf{k}} \left[\hat{a}_{\mathbf{k}\lambda}\boldsymbol{\varepsilon}_{\mathbf{k}\lambda} e^{-i(\omega_{\mathbf{k}}t-\boldsymbol{k}\cdot\boldsymbol{r})} + \mathrm{h.c.}\right], \tag{S1}$$

where $\omega_{\mathbf{k}} = c|\boldsymbol{k}|$, $\boldsymbol{\varepsilon}_{\mathbf{k}\lambda}$ are the polarization unit vectors orthogonal to $\boldsymbol{k}$, $\hat{a}_{\mathbf{k}\lambda}$ is the annihilation operator, h.c. is the Hermitian conjugate, and $N_{\mathbf{k}}$ is the normalization coefficient. The equal-time canonical commutator is equivalent to

$$\left[\hat{a}_{\mathbf{k}\lambda}, \hat{a}^{\dagger}_{\mathbf{k}'\lambda'}\right] = \delta_{\lambda\lambda'}\delta^{(3)}(\boldsymbol{k}-\boldsymbol{k}'), \qquad \sum_{\lambda=1,2} \boldsymbol{\varepsilon}_{\mathbf{k}\lambda,i}\,\boldsymbol{\varepsilon}^{*}_{\mathbf{k}\lambda,j} = \delta_{ij} - \frac{\boldsymbol{k}_i\boldsymbol{k}_j}{\boldsymbol{k}^2}. \tag{S2}$$

Requiring $\left[\widehat{\boldsymbol{A}}_i(\boldsymbol{r},t), \widehat{\boldsymbol{\Pi}}_j(\boldsymbol{r}',t)\right] = i\hbar\delta^{\perp}_{ij}(\boldsymbol{r}-\boldsymbol{r}')$ then fixes

$$N_{\mathbf{k}} = \sqrt{\frac{\hbar}{2\epsilon_0\omega_{\mathbf{k}}(2\pi)^3}}, \tag{S3}$$

which is Eq. (15) of the main text. The same expansion for field operators gives

$$\begin{aligned}
\widehat{\boldsymbol{E}}_{\perp}(\boldsymbol{r},t) &= i\int d^3k \sum_{\lambda=1,2} \omega_{\mathbf{k}} N_{\mathbf{k}}\left[\hat{a}_{\mathbf{k}\lambda}\boldsymbol{\varepsilon}_{\mathbf{k}\lambda} e^{-i(\omega_{\mathbf{k}}t-\boldsymbol{k}\cdot\boldsymbol{r})} - \mathrm{h.c.}\right],\\
\widehat{\boldsymbol{B}}(\boldsymbol{r},t) &= i\int d^3k \sum_{\lambda=1,2} N_{\mathbf{k}}\left[\hat{a}_{\mathbf{k}\lambda}(\boldsymbol{k}\times\boldsymbol{\varepsilon}_{\mathbf{k}\lambda}) e^{-i(\omega_{\mathbf{k}}t-\boldsymbol{k}\cdot\boldsymbol{r})} - \mathrm{h.c.}\right],\\
\widehat{H}_0 &= \int d^3k \sum_{\lambda=1,2} \hbar\omega_{\mathbf{k}}\left(\hat{a}^{\dagger}_{\mathbf{k}\lambda}\hat{a}_{\mathbf{k}\lambda} + \frac{1}{2}\right).
\end{aligned} \tag{S4}$$

The transverse sector is quantized canonically according to Eq. (S4), whereas the longitudinal component is fixed by the Gauss-law constraint without introducing an independent oscillator, as in Eqs. (10)–(13) of the main text, so the total quantized electric field is

$$\nabla^2\hat{\varphi}(\boldsymbol{r},t) = -\frac{\hat{\rho}(\boldsymbol{r},t)}{\epsilon_0}, \qquad \widehat{\boldsymbol{E}}(\boldsymbol{r},t) = -\frac{\partial\widehat{\boldsymbol{A}}(\boldsymbol{r},t)}{\partial t} - \nabla\hat{\varphi}(\boldsymbol{r},t). \tag{S5}$$

The Lorenz-gauge construction gives the same physical field. For a physical state $|\Psi\rangle$, the Gupta–Bleuler condition is

$$\left(\partial_\mu\hat{A}^\mu\right)^{(+)}|\Psi\rangle = 0, \qquad \left[\hat{a}_{(0)}(\boldsymbol{k}) - \hat{a}_{(3)}(\boldsymbol{k})\right]|\Psi\rangle = 0, \tag{S6}$$

and thus removes the scalar and longitudinal photon contributions from matrix elements of the gauge-invariant field tensor. Consequently,

$$\langle\Psi'|\hat{F}^{C}_{\mu\nu}|\Psi\rangle = \langle\Psi'|\hat{F}^{L}_{\mu\nu}|\Psi\rangle, \tag{S7}$$

where $\hat{F}^{C}_{\mu\nu}$ and $\hat{F}^{L}_{\mu\nu}$ are the Coulomb- and Lorenz-gauge field operators, respectively. These relations therefore establish the gauge-independent field content needed in the main text. Together, Eqs. (S1)–(S7) fix the normalized transverse photon operator as the independent radiative degree of freedom, while the longitudinal electric field remains constrained by the aperture charge. This separation provides the canonical basis for the source coupling considered next.

## S2. SOURCE COUPLING AND COHERENT DISPLACEMENT

Having fixed the physical photon basis, we next determine how the localized Bethe source interacts with it. At this stage the aperture is represented only by its effective magnetic and electric moments; its finite transverse geometry is not yet invoked. Their interaction with the transverse field is given by Eq. (21) of the main text:

$$\widehat{H}_{\mathrm{int}}(t) = -\boldsymbol{m}(t)\cdot\left(\nabla\times\widehat{\boldsymbol{A}}\right)|_{\boldsymbol{r}=0} + \boldsymbol{p}(t)\cdot\frac{\partial\widehat{\boldsymbol{A}}}{\partial t}|_{\boldsymbol{r}=0}. \tag{S8}$$

For a radiative mode with $\boldsymbol{n} = \frac{\boldsymbol{k}}{|\boldsymbol{k}|}$, the electric moment enters through the transverse projection used in the main text,

$$\boldsymbol{p}_{0\perp} = \boldsymbol{n} \times (\boldsymbol{p}(t) \times \boldsymbol{n}). \tag{S9}$$

The single-mode matrix element then becomes

$$V_{3D}(\boldsymbol{k}, \lambda) = -iN_{\mathbf{k}}\boldsymbol{\varepsilon}^{*}_{\mathbf{k}\lambda} \cdot (\boldsymbol{k} \times \boldsymbol{m}_0 - \omega_{\mathbf{k}}\boldsymbol{p}_{0\perp}), \tag{S10}$$

which reproduces Eq. (28) of the main text. For a monochromatic source switched on as $e^{\eta t}$, $\eta \to 0^+$, the resonant part of the Heisenberg equation is

$$\frac{d}{dt}\left[\hat{a}_{\mathbf{k}\lambda}(t)e^{i\omega_{\mathbf{k}}t}\right] = -\frac{i}{\hbar}V_{3D}(\boldsymbol{k}, \lambda)e^{i(\omega_{\mathbf{k}}-\omega_0-i\eta)t}. \tag{S11}$$

Retaining only the driven contribution and solving Eq. (S11) gives

$$\hat{a}_{\mathbf{k}\lambda}(t) = -\frac{V_{3D}(\boldsymbol{k}, \lambda)}{\hbar}\frac{e^{-i\omega_0 t+\eta t}}{\omega_{\mathbf{k}} - \omega_0 - i\eta}. \tag{S12}$$

The corresponding steady-state coherent amplitude is therefore

$$\alpha_{3D}(\boldsymbol{k}, \lambda) = \langle \hat{a}_{\mathbf{k}\lambda}(t) \rangle = -\frac{V_{3D}(\boldsymbol{k}, \lambda)}{\hbar}\frac{1}{\omega_{\mathbf{k}} - \omega_0 - i\eta}, \tag{S13}$$

in agreement with Eq. (27) of the main text. Substitution into Eq. (S1) gives the three-dimensional coherent field,

$$\langle \widehat{\boldsymbol{A}}(\boldsymbol{r}, t) \rangle = \int d^2k_\perp \int dk_z \sum_\lambda \alpha_{3D}\,(\boldsymbol{k}, \lambda)N_{\mathbf{k}}\boldsymbol{\varepsilon}_{\mathbf{k}\lambda}e^{-i(\omega_{\mathbf{k}}t-\boldsymbol{k}\cdot\boldsymbol{r})}. \tag{S14}$$

At fixed $\boldsymbol{k}_\perp$, the change of variables used in the main text is

$$dk_z = d\omega_{\mathbf{k}}\frac{\omega_{\mathbf{k}}}{c^2k_z}, \qquad \langle \widehat{\boldsymbol{A}}(\boldsymbol{r}, t) \rangle = \int d^2k_\perp \sum_\lambda [-\frac{2\pi i\omega_0}{\hbar c^2 k_z}N_{\mathbf{k}}V_{3D}(\boldsymbol{k}, \lambda)]\,\boldsymbol{\varepsilon}_{\mathbf{k}\lambda}e^{-i(\omega_0 t-\boldsymbol{k}\cdot\boldsymbol{r})}. \tag{S15}$$

Equation (S15) is the two-dimensional angular-spectrum form of Eq. (29) of the main text. It shows that the aperture field is generated by the coherent displacement of the normalized free-space photon basis. This completes the source part of the construction and leaves the transverse source distribution as the only missing ingredient of the finite-aperture field.

## S3. FINITE APERTURE AND THE COMPLETE WEYL FIELD

The Weyl amplitude is now fixed apart from the finite transverse extent of the source. We therefore isolate the geometric contribution in this section and require it to preserve both the point-aperture limit and the treatment of propagating and evanescent components used in the main text. The propagating and evanescent components can be written as:

$$\boldsymbol{k}_{\mathbf{z}} = \sqrt{\boldsymbol{k}_{\mathbf{0}}^2 - \boldsymbol{k}_\perp^2} = \begin{cases} \sqrt{\boldsymbol{k}_{\mathbf{0}}^2 - \boldsymbol{k}_\perp^2}, & \boldsymbol{k}_\perp < \boldsymbol{k}_{\mathbf{0}}, \\ i\sqrt{\boldsymbol{k}_\perp^2 - \boldsymbol{k}_{\mathbf{0}}^2}, & \boldsymbol{k}_\perp > \boldsymbol{k}_{\mathbf{0}}, \end{cases} \qquad \boldsymbol{k}_{\mathbf{0}} = \frac{\omega_0}{c}, \tag{S16}$$

where the evanescent branch is chosen with $\operatorname{Im} k_z \geq 0$. Using Eq. (S10), the polarization completeness relation in Eq. (S2), and $c^{-2} = \mu_0\epsilon_0$, Eq. (S15) reduces to the point-source result

$$\langle \widehat{\boldsymbol{A}}(\boldsymbol{r}, t) \rangle_{\text{pt}} = -\frac{\mu_0}{8\pi^2}\int d^2k_\perp \frac{\boldsymbol{k} \times \boldsymbol{m}_0 - \omega_0\boldsymbol{p}_{0\perp}}{k_z}e^{-i(\omega_0 t-\boldsymbol{k}_x x-\boldsymbol{k}_y y-\boldsymbol{k}_z z)}. \tag{S17}$$

Finite aperture size enters only through the in-plane source distribution. For a circular aperture of radius $a$,

$$\widetilde{\boldsymbol{m}}(\boldsymbol{\varrho}) = \boldsymbol{m}_0 f(\boldsymbol{\varrho}), \qquad \widetilde{\boldsymbol{p}}(\boldsymbol{\varrho}) = \boldsymbol{p}_{0\perp}f(\boldsymbol{\varrho}), \tag{S18}$$

with $\boldsymbol{\varrho} = |\boldsymbol{\varrho}| \leq a$ and

$$f(\boldsymbol{\varrho}) = \frac{3}{2\pi a^2}\sqrt{1 - \left(\frac{\boldsymbol{\varrho}}{a}\right)^2}. \tag{S19}$$

The transverse Fourier transform gives

$$f(\boldsymbol{k}_\perp) = \int d^2\boldsymbol{\varrho}\, f(\boldsymbol{\varrho}) e^{-i\mathbf{k}_\perp\cdot\boldsymbol{\varrho}} = 3\int_0^1 \boldsymbol{u}\, du\sqrt{1-\boldsymbol{u}^2}\, J_0(\boldsymbol{k}_\perp a\boldsymbol{u}), \tag{S20}$$

where $\boldsymbol{u} = \boldsymbol{\varrho}/a$ and $J_0$ is the Bessel function of the first kind of order zero (cylindrical Bessel function). Applying Sonine's finite integral yields the normalized aperture spectrum

$$\int_0^1 \boldsymbol{u}\, du\sqrt{1-\boldsymbol{u}^2}\, J_0(c\boldsymbol{u}) = \frac{j_1(c)}{c}, \qquad f(\boldsymbol{k}_\perp) = \frac{3j_1(\boldsymbol{k}_\perp a)}{\boldsymbol{k}_\perp a}. \tag{S21}$$

Because $\frac{3j_1(x)}{x} \to 1$ as $x \to 0$, the total Bethe moments are preserved. Their spectra are consequently

$$\tilde{\boldsymbol{m}}(\boldsymbol{k}_\perp) = \boldsymbol{m}_0 \frac{3j_1(\boldsymbol{k}_\perp a)}{\boldsymbol{k}_\perp a}, \qquad \tilde{\boldsymbol{p}}(\mathbf{k}_\perp) = \boldsymbol{p}_{0\perp}\frac{3j_1(\boldsymbol{k}_\perp a)}{\boldsymbol{k}_\perp a}. \tag{S22}$$

The finite-aperture coupling in Eq. (41) is thus

$$V_{\mathbf{k}_\perp,\lambda}^{(\text{hole})} = -iN_{\mathbf{k}_\perp}\boldsymbol{\varepsilon}_{\mathbf{k}_\perp\lambda}^* \cdot [\boldsymbol{k}\times\tilde{\boldsymbol{m}}(\boldsymbol{k}_\perp) - \omega_0\tilde{\boldsymbol{p}}(\boldsymbol{k}_\perp)]. \tag{S23}$$

Substituting Eq. (S22) into Eq. (S17) gives

$$\langle\hat{\boldsymbol{A}}(\boldsymbol{r},t)\rangle = -\frac{\mu_0}{8\pi^2}\int d^2k_\perp\, \frac{\boldsymbol{k}\times\boldsymbol{m}_0 - \omega_0\boldsymbol{p}_{0\perp}}{\boldsymbol{k}_{\mathbf{z}}}\cdot\frac{3j_1(\boldsymbol{k}_\perp a)}{\boldsymbol{k}_\perp a} e^{-i(\omega_0 t - \boldsymbol{k}_\perp\cdot\boldsymbol{\varrho} - \boldsymbol{k}_z z)}. \tag{S24}$$

Equation (S24) is precisely Eq. (42) of the main text. The normalized Bessel factor contains the entire finite-radius correction, while the common branch of $k_z$ retains both the propagating sector ($k_\perp < k_0$) and the evanescent sector ($k_\perp > k_0$) in a single expression. Thus, the finite aperture modifies the angular-spectrum weight without altering either the canonical normalization or the source projection established above. Equation (S24) is consequently the complete transverse field from which the observable fields follow.

## S4. OBSERVABLE FIELDS AND THE BETHE LIMIT

With the transverse expectation value fixed by Eq. (42), the complete electric field is obtained by restoring the constrained longitudinal contribution. The same construction must also reproduce the Bethe transmission law in the far-field limit. The observable magnetic and electric fields are first obtained as in Eqs. (43)–(45):

$$\langle\hat{\boldsymbol{B}}(\boldsymbol{r},t)\rangle = \nabla\times\langle\hat{\boldsymbol{A}}(\boldsymbol{r},t)\rangle, \qquad \langle\hat{\boldsymbol{E}}(\boldsymbol{r},t)\rangle = i\omega_0\langle\hat{\boldsymbol{A}}(\boldsymbol{r},t)\rangle - \nabla\langle\hat{\varphi}(\boldsymbol{r},t)\rangle. \tag{S25}$$

The scalar potential contains no independent photon oscillator; it is determined by the Bethe polarization charge as in Eq. (45),

$$\langle\hat{\varphi}(\mathbf{r},t)\rangle = \frac{1}{4\pi\epsilon_0}\int d^3r'\, \frac{\langle\hat{\rho}(\boldsymbol{r}',t)\rangle}{|\boldsymbol{r}-\boldsymbol{r}'|}. \tag{S26}$$

In the far field, the evanescent sector is exponentially suppressed, and the aperture factor tends to unity. Integration of the transmitted dipole radiation gives Eq. (46),

$$P_{\text{rad}} = \frac{\mu_0 c\boldsymbol{k}_0^4}{6\pi}(|\boldsymbol{m}_0|^2 + c^2\boldsymbol{p}_0^2), \tag{S27}$$

while the incident energy flux in Eq. (47) is

$$S_{\text{inc}} = \frac{1}{2}\mu_0 c\boldsymbol{H}_0^2 = \frac{1}{2Z_0}\boldsymbol{E}_0^2. \tag{S28}$$

Using the Bethe moments of Eqs. (3) and (4) of the main text, we obtain

$$\sigma_t = \frac{P_{\text{rad}}}{S_{\text{inc}}} = \frac{64}{27\pi}\boldsymbol{k}_0^4 a^6, \tag{S29}$$

which is Eq. (48) of the main text. Equations (S25)–(S29) complete the connection between the quantized source construction and its measurable field content. The transverse Weyl field of Eq. (42) retains the finite-aperture propagating and evanescent spectrum, the charge-determined scalar potential restores the longitudinal electric field, and the coherent far-field limit recovers Bethe's transmission law without an additional phenomenological normalization. The near-field and far-field results are therefore two limits of the same source-defined quantum aperture field.